\documentclass[12pt]{article}
\usepackage{graphicx}
\usepackage{amsmath}
\usepackage{amssymb}
\usepackage{caption2}
\begin{document}
\bibliographystyle{prsty}
\begin{center}
{\large {\bf \sc{ Analysis of the $X(1576)$
as a tetraquark state with  the QCD sum rules }}} \\[2mm]
Zhi-Gang Wang$^{1}$ \footnote{Corresponding author;
E-mail,wangzgyiti@yahoo.com.cn. }, Shao-Long Wan$^{2}$  \\
$^{1}$ Department of Physics, North China Electric Power University, Baoding 071003, P. R. China \\
$^{2}$ Department of Modern Physics, University of Science and Technology of China, Hefei 230026, P. R. China \\
\end{center}

\begin{abstract}
In this letter, we take the point of view that the $X(1576)$ be
tetraquark state which consists of  a scalar-diquark and an
anti-scalar-diquark in relative $P$-wave,  and calculate its mass in
the framework of the QCD sum rules approach.  The numerical value of
the mass  $m_X=(1.66\pm 0.14) GeV$ is consistent with the
experimental data, there may be some tetraquark component in the
vector meson $X(1576)$.
\end{abstract}


\bigskip

The BES collaboration observed  a broad resonance  $X(1576)$ in the
$K^+K^-$ invariant mass spectrum in  the decay $J/\psi\to
K^+K^-\pi^0$, the pole position is
$1576^{+49+98}_{-55-91}-i409^{+11+32}_{-12-67}$ $ MeV$ and the
$J^{PC}$ is $ 1^{--}$ \cite{BES06}. The resonance $X(1576)$
 is  an isovector due to the $\pi$ produced from the isoscalar
$J/\psi$ \cite{GS06,DY06,KL06}. The broad resonance cannot be
interpreted  as any known mesons or their mixing states, it may be a
tetraquark state or a meson-meson molecular state, in this letter,
we study the possibility of the tetraquark state assignment. In the
tetraquark scenario, the $X(1576)$ consists of the basic
constituents, the scalar diquarks $[us]$, $[ds]$, $[\bar{u}\bar{s}]$
and $[\bar{d}\bar{s}]$, the decay $X(1576) \to K^{+}K^{-}$ occurs
through  the fall apart mechanism with re-arrangement  in the color
space, the width of the $X(1576)$ can be much larger than the width
of any known two-quark vector mesons \cite{GS06,DY06,KL06}.

It is obvious that the vector tetraquark state has many novel
features which are of phenomenological importance and have many
observable  consequence, one can consult Refs.\cite{GS06,DY06,KL06}
for details. In this letter, we focus on its substructure and mass,
and take the point of view that the vector  meson $X(1576)$ with the
isospin $I=1$ be a tetraquark state,  calculate the value of its
mass in the framework of the QCD sum rules approach
\cite{Shifman79}.

In the following, we write down the interpolating current  for the
vector  meson $X(1576)$,
\begin{eqnarray}
J_\mu(x)&=& D^a (x)\partial_\mu \bar{D}^a(x)-\partial_\mu D^a(x)\bar{D}^a (x)\nonumber \\
&&-U^a (x)\partial_\mu \bar{U}^a(x)+\partial_\mu U^a(x)  \bar{U}^a (x), \\
U^a(x)&=&\epsilon^{abc}d_b^T(x)C\gamma_5s_c(x) , \nonumber \\
D^a(x)&=&\epsilon^{abc}u_b^T(x)C\gamma_5s_c (x), \nonumber
\end{eqnarray}
where the $a,~b,~c$ are color indices and the $C$ is the charge
conjugation matrix.  The constituents $U^a(x)$ and $D^a(x)$
represent the scalar diquarks ($S^a$) with the $J^P=0^+$, they
belong to the antitriplet $\bar{3}_c$ representation of the color
$SU(3)$ group and can cluster together to form the scalar-scalar
($S^a-\bar{S}^a$) type diquark  pairs to give the correct spin and
parity of the vector mesons $J^{P}=1^{-}$. The attractive
interaction of one-gluon exchange favors  the formation of diquarks
in the color antitriplet $\overline{3}_{ c}$, flavor antitriplet
$\overline{3}_{ f}$ and spin singlet  $ 1_{s} $. The instanton
induced force results in strong attraction in the scalar diquark
channels and strong repulsion in the pseudoscalar diquark ($P^a$)
channels, we prefer the $S^a-\bar{S}^a$ type interpolating currents
to the $P^a-\bar{P}^a$ type interpolating currents if the instantons
manifest themselves \cite{GluonInstanton,WYW05}.   The strong
attractions between the scalar diquark states $S^a$ and $\bar{S}^a$
in $S$-wave may result in a nonet manifested below $1GeV$,  while
the conventional $^3P_0$ $\bar{q}q$ nonet would have masses about
$1.2-1.6 GeV$ \cite{WYW05,ReviewScalar}. The additional $P$-wave
between the $S^a$ and $\bar{S}^a$  can lead to the negative parity
and much higher mass than $1GeV$, for example, the contribution from
the $P$-wave is about $480MeV$ in the Isgur-Karl model \cite{Isgur},
we introduce the relative $P$-wave through the $\partial_\mu$
between the $S^a$ and $\bar{S}^a$. If we take the scalar diquark
mass $m_D=0.7GeV$ from lattice QCD as the input parameter
\cite{LattDiquark}, the mass of vector tetraquark is about
$2m_D+0.48=1.88GeV$, our numerical result $m_X=(1.66\pm 0.14) GeV$
confirms  this naive analysis. In fact, there are large
uncertainties about the masses  of the colored diquarks, for more
literatures and detailed discussions about this subject, one can
consult Ref.\cite{WangDiquark}.

In this letter, we study the mass of the $X(1576)$ with  the
following two-point correlation function,
\begin{eqnarray}
\Pi_{\mu \nu}(p)=i\int d^4x e^{ip \cdot x}\langle 0
|T\{J_\mu(x)J_\nu^\dagger(0)\}|0\rangle.
\end{eqnarray}
The correlation function $\Pi_{\mu\nu}(p)$ can be decomposed as
\begin{eqnarray}
\Pi_{\mu\nu}&=&-\Pi \left\{g_{\mu\nu}-\frac{p_\mu
p_\nu}{p^2}\right\}+\Pi_0\frac{p_\mu p_\nu}{p^2},
\end{eqnarray}
due to the Lorentz covariance. The transverse  part $\Pi$ comes from
the contributions of  the vector mesons and the longitudinal part
$\Pi_0$ comes from the scalar mesons.  We derive the sum rules with
the tensor structure $ \left\{g_{\mu\nu}-\frac{p_\mu
p_\nu}{p^2}\right\}$, for the scalar meson, one can use the scalar
interpolating currents as the ideal choice rather than the vector
current $J_\mu(x)$ \cite{WYW05}.

 According to
the basic assumption of the current-hadron duality in the QCD sum
rules approach \cite{Shifman79}, we insert  a complete series of
intermediate states  satisfying the unitarity   principle with the
same quantum numbers as the current operator $J_\mu(x)$
 into the correlation function in
Eq.(2)  to obtain the hadronic representation. Isolating the ground
state contribution from the pole term of the vector tetraquark
state, we obtain the result,
\begin{eqnarray}
\Pi(p)=\frac{2f_X^2}{m_X^{2}-p^2}+\cdots \, ,
\end{eqnarray}
where the following definition has  been used,
\begin{equation}
 \langle 0 | J_\mu(0)|X\rangle =\sqrt{2}f_X  \epsilon_\mu ,
 \end{equation}
 the $\epsilon_\mu$ is the polarization vector of the $X(1576)$.
We have not shown the contributions from the higher resonances and
continuum states explicitly for simplicity.

The  calculation of the operator product expansion in the  deep
Euclidean space-time region is
  straightforward and tedious, technical details are neglected for
  simplicity.  In this letter, we consider the terms of the vacuum condensates add up to dimension 11,
  neglect the terms proportional to $m_s^2$ , $\langle \bar{q}g_s \sigma G q\rangle^2$ ,
  $\langle \bar{s}g_s \sigma G s\rangle^2$ and
   $\langle \bar{q}g_s \sigma G q\rangle\langle \bar{s}g_s \sigma G s\rangle$.
   In calculation, we
 take the assumption of the vacuum saturation for the high
dimension vacuum condensates, they  are always
 factorized to lower condensates with the vacuum saturation in the QCD sum rules approach,
 the factorization works well in the large $N_c$ limit.
  Once  the analytical  result is obtained,
  then we can take the current-hadron duality below the threshold
$s_0$ and perform the Borel transformation with respect to the
variable $P^2=-p^2$, finally we obtain  the following sum rule,
\begin{eqnarray}
f^{2}_{X}e^{-\frac{m_{X}^{2}}{M^2}}=AA ,
\end{eqnarray}
\begin{eqnarray}
AA&=&\int_{4m_s^2}^{s_0}dt e^{-\frac{t}{M^2}}\left\{
\frac{t^5}{716800 \pi^6} +\frac{\langle \bar{s}s \rangle \langle
\bar{q}g_s \sigma G q\rangle+\langle \bar{q}q \rangle\langle
\bar{s}g_s \sigma  G s\rangle}{24 \pi^2} t \right. \nonumber\\
&&\left.-\frac{2\langle \bar{q} q\rangle-\langle \bar{s}
s\rangle}{1920 \pi^4}m_s t^3 +\frac{t^3}{15360 \pi^4} \langle
\frac{\alpha_s GG}{\pi}\rangle \right\} +  \nonumber\\
&&  \frac{\langle \bar{s}s \rangle^2 \langle \bar{q}g_s \sigma G
q\rangle+\langle \bar{s}s \rangle \langle \bar{q}q \rangle \langle
\bar{s}g_s \sigma G s\rangle-2\langle \bar{s}s \rangle \langle
\bar{q}q \rangle \langle \bar{q}g_s \sigma G q\rangle-2\langle
\bar{q}q \rangle^2 \langle \bar{s}g_s \sigma G s\rangle}{18}m_s ,
\nonumber
\end{eqnarray}
here the   $s_0$ is the threshold parameter.
 Differentiate the above sum rule with respect to the
variable $\frac{1}{M^2}$, then eliminate the quantity $f_{X}$,
  we obtain
\begin{eqnarray}
m^{2}_{X}&=&BB/AA , \\
BB&=&\int_{4m_s^2}^{s_0}dt e^{-\frac{t}{M^2}}\left\{
\frac{t^6}{716800 \pi^6} +\frac{\langle \bar{s}s \rangle \langle
\bar{q}g_s \sigma G q\rangle+\langle \bar{q}q \rangle\langle
\bar{s}g_s \sigma  G s\rangle}{24 \pi^2} t^2 \right. \nonumber\\
&&\left.-\frac{2\langle \bar{q} q\rangle-\langle \bar{s}
s\rangle}{1920 \pi^4}m_s t^4 +\frac{t^4}{15360 \pi^4} \langle
\frac{\alpha_s GG}{\pi}\rangle \right\}  .\nonumber
\end{eqnarray}
 It is easy to perform the   $t$ integral in
 Eqs.(6-7),  we prefer this form for simplicity.

The input parameters are taken as $\langle \bar{s}s \rangle=(0.8\pm
0.1)\langle \bar{q}q\rangle$, $\langle \bar{s}g_s\sigma G s
\rangle=m_0^2\langle \bar{s}s \rangle$, $\langle \bar{q}g_s\sigma
 G q \rangle=m_0^2\langle \bar{q}q \rangle$, $m_0^2=(0.8\pm0.1)GeV^2$,  $\langle \bar{q}q \rangle=-(0.24 \pm
0.01 GeV)^3$, $\langle \frac{\alpha_sGG}{\pi} \rangle=(0.33 GeV)^4$,
 $m_u=m_d=0$ and $m_s=(0.14\pm 0.01)GeV$.
  The threshold parameter is taken as $\sqrt{s_0}=(2.1\pm0.1)GeV$
 to avoid possible contaminations
 from the  high resonances and continuum states. The width of the $X(1576)$
 is  very broad, about $0.8GeV$, the threshold parameter $s_0\geq (1.6+0.4)$ can include the
  contribution from the ground state vector meson. The Borel parameter $M^2$ is chosen to be
  $ M^2=(2-5)GeV^2$, i.e. $M=(1.4-2.2)GeV$ and $e^{-\frac{s_0}{M^2}}\leq e^{-1}$,
  the contributions from the high resonances and continuum states can be suppressed, furthermore,
   in this region, the mass from the sum rule is
  rather stable, which is shown in Fig.1.  For detailed
  discussions about the  criterion  of choosing the Borel parameter $M^2$
  and threshold parameter $s_0$ for the multiquark states, one can consult
  Ref.\cite{WangWan06}. The contribution from the perturbative term is about $10\%$, the dominating contribution comes from the term
    $\langle \bar{s}s \rangle \langle
\bar{q}g_s \sigma G q\rangle+\langle \bar{q}q \rangle\langle
\bar{s}g_s \sigma  G s\rangle$, about $80\%$, the other terms are of
minor importance,  we can expect the convergence of the operator
product expansion.

\begin{figure}
 \centering
 \includegraphics[totalheight=7cm,width=10cm]{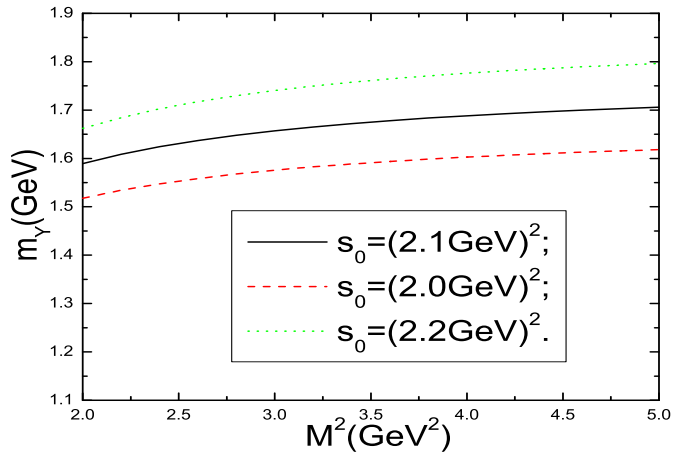}
  \caption{The   $m_X$  with the Borel parameter $M^2$ for the central values of the condensates. }
\end{figure}

Taking into account all the uncertainties, finally we obtain the
mass  of the $X(1576)$,
  \begin{eqnarray}
  m_X=(1.66\pm 0.14) GeV.
  \end{eqnarray}

Comparing with the experimental data ($1576^{+49+98}_{-55-91}MeV$)
and the calculation of the quark model ($1632.854MeV$), our
numerical result is reasonable, although somewhat larger, the vector
meson $X(1576)$ may have some tetraquark component. In this letter,
we take the scalar diquarks as the basic constituents, the
instantons manifest themselves in  the scalar diquark channels, the
contributions of the direct instantons may account for  the
discrepancy \cite{Forkel}.

The tetraquark state can mix with the molecular state if they have
the same quantum numbers,  the mixing can also pull  the mass down.
In Ref.\cite{GS06}, the $X(1576)$ is taken as an  $S$-wave
$K^*(892)$-${\bar \kappa}$ molecular state,  the large width of the
$X(1576)$ is taken into account by the  large width of the $\kappa$.
If the negative bound energy of the molecular state is large enough,
the mixing can pull the mass down significantly. In  the tetraquark
scenario, the dominant decay modes of the $ X(1576)$ are
$K^{+}K^{-}$, $K_{L}K_{S}$ and $\phi\pi^{0}$ which are OZI-allowed,
not the  $\pi^{+}\pi^{-}$ which is OZI-forbidden \cite{DY06}. In the
$K^*(892)$-${\bar \kappa}$ molecular scenario, the decay mode $
\pi^+\pi^-$ has much larger branching ratio than the decay mode $
K^+K^-$  \cite{GS06}. Whether or not there exist the resonant
structure in the decay $J/\psi\to \pi^+\pi^-\pi^0$ is of great
importance for understanding the structure of the $X(1576)$.

In this letter, we take the point of view that the $X(1576)$ be
tetraquark state which consists  of  a scalar-diquark and an
anti-scalar-diquark in relative $P$-wave,  and calculate its mass in
the framework of the QCD sum rules approach.  The numerical value of
the mass  is consistent with the experimental data, there may be
some tetraquark component in the vector meson $X(1576)$.

\section*{Acknowledgment}
This  work is supported by National Natural Science Foundation,
Grant Number 10405009,  and Key Program Foundation of NCEPU.

\end{document}